\documentclass[conference]{IEEEtran}
\IEEEoverridecommandlockouts

\usepackage{colortbl}
\usepackage[table]{xcolor}
\usepackage{amsmath,amssymb,amsfonts}
\usepackage{algorithm}
\usepackage{algpseudocode}
\usepackage{graphicx}
\usepackage{siunitx}
\usepackage{textcomp}
\usepackage{multirow}
\usepackage{arydshln}
\usepackage{caption}
\usepackage{subcaption}
\usepackage{multirow}
\usepackage{rotating}
\usepackage{booktabs}
\usepackage[sortcites, sorting=none, style=numeric-comp]{biblatex}
\usepackage{acro}
\usepackage{dblfloatfix}
\usepackage{subcaption}
\usepackage{wrapfig}

\ifCLASSOPTIONcompsoc
\else
\fi

\usepackage{tikz}

\usepackage{hyperref}
\usepackage{cleveref}

\def\BibTeX{{\rm B\kern-.05em{\sc i\kern-.025em b}\kern-.08em T\kern-.1667em\lower.7ex\hbox{E}\kern-.125emX}}

\makeatletter
\let\old@ps@headings\ps@headings
\let\old@ps@IEEEtitlepagestyle\ps@IEEEtitlepagestyle
\def\confheader#1{%
\def\ps@headings{%
\old@ps@headings%
\def\@oddhead{\strut\hfill#1\hfill\strut}%
\def\@evenhead{\strut\hfill#1\hfill\strut}%
}%
\def\ps@IEEEtitlepagestyle{%
\old@ps@IEEEtitlepagestyle%
\def\@oddhead{\strut\hfill#1\hfill\strut}%
\def\@evenhead{\strut\hfill#1\hfill\strut}%
}%
\ps@headings%
}
\makeatother

\title{ChronoChart: Setup-Agnostic Channel Charting for Label-Free Indoor Localization

\thanks{The authors would like to thank the Federal Ministry of Research, Technology, and Space (BMFTR) for supporting the xG‑RIC project as part of the research program Communication Systems “Souverän. Digital. Vernetzt.” (grant number 16KIS2429K).}}

\def\authorrefmark#1{\ensuremath{^{\textbf{#1}}}}
\author{
    Majid Sharghi Foroushani\authorrefmark{1},
    Jonas Pirkl\authorrefmark{1},
    Jonathan Ott\authorrefmark{1},
    Maximilian Stahlke\authorrefmark{1},
    George Yammine\authorrefmark{1},
    Tobias Feigl\authorrefmark{1,}\authorrefmark{2},\\
    and Christopher Mutschler\authorrefmark{1,}\authorrefmark{3}\\
    {\tt\footnotesize\{firstname.lastname\}@iis.fraunhofer.de}\\ 
    \IEEEauthorblockA{\authorrefmark{1}
        Fraunhofer Institute for Integrated Circuits (IIS), 
        Division Positioning and Networks, 
        90411 Nürnberg, Germany
    }
    \IEEEauthorblockA{\authorrefmark{2}
        Programming Systems Lab,
        Friedrich-Alexander-Universität Erlangen-Nürnberg, 
        91052 Erlangen, Germany 
    }
    \IEEEauthorblockA{\authorrefmark{3}
        Machine Learning and Positioning Systems Lab,
        University of Technology Nuremberg (UTN), 
        Nürnberg, Germany
    }
}

\usepackage[acronym]{glossaries}
\glsdisablehyper
\newacronym{gnn}{GNN}{Graph Neural Network}
\newacronym{vae}{VAE}{Variational Autoencoder}
\newacronym{ae}{AE}{autoencoder}
\newacronym{ml}{ML}{Machine Learning}
\newacronym{cc}{CC}{Channel Charting}
\newacronym{csi}{CSI}{Channel State Information}
\newacronym{cir}{CIR}{Channel Impulse Response}
\newacronym{nlos}{NLOS}{non-line-of-sight}
\newacronym{los}{LOS}{line-of-sight}
\newacronym{mlp}{MLP}{multi-layer perceptron}
\newacronym{cnn}{CNN}{convolutional neural network}

\begin{document}

\maketitle

\begin{abstract}
Accurate radio-based indoor localization remains challenging in multipath-rich and dynamic indoor environments. 
Supervised fingerprinting methods achieve high positioning accuracy by learning mappings from radio channel measurements to physical locations, but require costly position labels and frequent recalibration. 
\gls{cc} addresses this limitation by learning low-dimensional representations of \gls{csi} without position labels, yielding channel charts in which nearby points correspond to nearby physical locations. 
However, prominent \gls{cc} methods rely on setup-specific dissimilarity metrics, such as ADP-, CIRA-, or PDR-based distances, and on precomputed $O(n^2)$ distance matrices, limiting scalability and transferability across radio systems.
We propose ChronoChart, a timestamp-supervised \gls{cc} framework that requires only \gls{csi} and timestamps without prior knowledge for training. 
Our method jointly optimizes complementary objectives: a triplet loss using temporally derived positive and negative samples to establish coarse global structure, a temporal-proximity loss that promotes smooth trajectories in the chart, and a joint-embedding loss that enforces consistency between representations of different augmentations. 
Evaluated on real-world 5G, MIMO, and passive UWB data, our method matches or exceeds prior \gls{cc} baselines that depend on tailored distance metrics, and remains scalable and applicable across heterogeneous radio systems.
\end{abstract}

\vspace{-0.2cm}

\section{Introduction}

Radio-based indoor localization is a key enabler in low-visibility environments~\cite{ref1}. 
However, accurate localization remains challenging, as reflections, shadowing, blocked \gls{los} paths, and environmental dynamics strongly affect signal propagation. 
Traditional RSSI-, AoA-, or TDoA-based methods therefore depend on accurate propagation models and often degrade in complex indoor environments. 
Fingerprinting methods achieve high positioning accuracy by matching observed \gls{csi} to labeled reference measurements~\cite{ref2,ref3}, but require extensive site surveys and frequent recalibration when the environment changes. 
This limits their practicality in large-scale or dynamic deployments.
\gls{cc} has emerged as a label-free alternative to supervised fingerprinting. 
It embeds high-dimensional \gls{csi} into a low-dimensional channel chart such that nearby chart points correspond to nearby physical locations~\cite{ref4,ref6}. 
As training does not require ground-truth coordinates, \gls{cc} substantially reduces deployment effort. 
Manifold-learning methods (PCA, Isomap, Sammon mapping) reveal local structure, but lack a learnable mapping for unseen samples~\cite{ref13}. 
Parametric approaches, i.e., autoencoders~\cite{ref4}, Siamese networks~\cite{ref7}, and triplet networks~\cite{triplet}, address this limitation by learning encoders that are applied to new measurements.

\begin{figure}[t]
    \centering
    \includegraphics[width=0.70\linewidth]{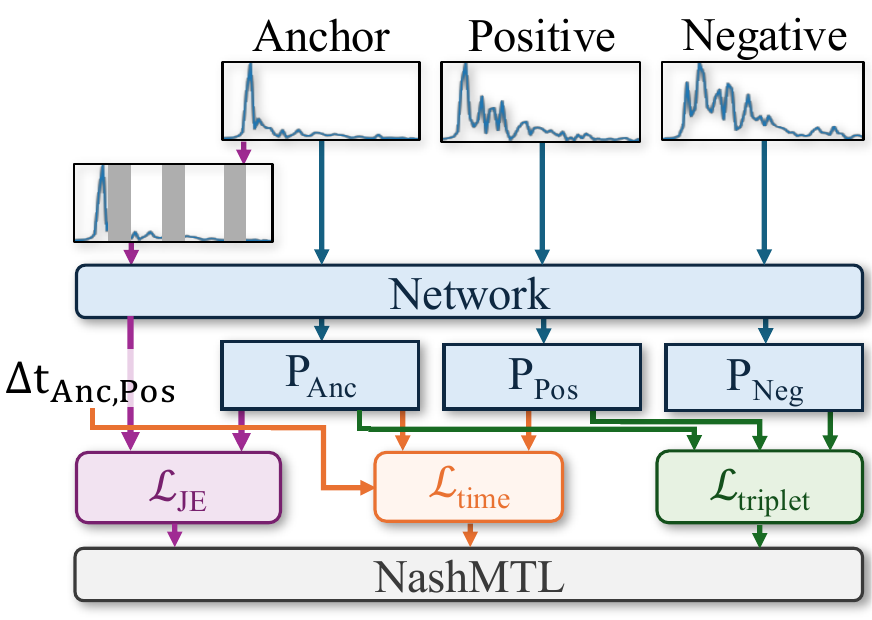}
    \vspace{-0.2cm}
    \caption{ChronoChart trains a shared encoder from CSI samples and timestamps using temporally derived triplets. A triplet loss (green), temporal-proximity loss (orange), and joint-embedding loss (purple) are combined with NashMTL, without setup-specific priors.}
    \label{fig:overview}
    \vspace{-0.6cm}
\end{figure}

Despite this progress, prominent \gls{cc} methods rely on setup-specific design choices. 
Geodesic-distance-based Siamese methods improve global consistency, but require precomputed pairwise distance matrices with $O(n^2)$ memory complexity~\cite{geodesic}. 
They also depend on radio-specific dissimilarity metrics, such as ADP-, CIRA-, or PDR-based distances~\cite{adp,geodesic,pdr}. 
Others incorporate temporal, inertial, or acceleration constraints to improve trajectory smoothness and global chart structure~\cite{ae-cc,rappaport2021improving,uncertainty-aware}. 
While effective, these approaches require additional priors, auxiliary sensors, carefully designed metrics, or expensive preprocessing. 
Thus, an open problem is how to obtain globally coherent channel charts without position labels, radio-specific distance metrics, propagation priors, floor maps, inertial measurements, or precomputed distance matrices.

To address this problem, we introduce \emph{ChronoChart}, a timestamp-supervised \gls{cc} framework for label-free indoor localization. 
ChronoChart replaces setup-specific distance engineering with weak supervision from timestamps: samples that are close in time are likely to be spatially close, while temporally distant samples provide contrastive structure. 
As illustrated in Fig.~\ref{fig:overview}, it trains a shared encoder from \gls{csi} and timestamps only by jointly optimizing three complementary objectives: a triplet loss with temporally derived positives and negatives, a temporal-proximity loss for smooth chart trajectories, and a joint-embedding loss for robust data-driven representations. 
The objectives are combined using~\cite{nashmtl} to mitigate gradient conflicts during multi-objective training. 
Thus, ChronoChart eliminates the need of auxiliary information.
We evaluate ChronoChart across heterogeneous real-world radio systems, including Fraunhofer 5G, Dichasus distributed MIMO, and passive UWB. 
Experiments show that ChronoChart outperforms non-parametric \gls{cc}, triplet-only training, and Siamese baselines on the Fraunhofer 5G and passive UWB datasets, while achieving performance comparable to geodesic-loss methods on Dichasus. 
Ablation studies further confirm that the triplet, temporal-proximity, and joint-embedding objectives provide complementary improvements and improve robustness under non-uniform motion.

The paper is organized as follows. 
Sec.~\ref{sec:related_work} reviews related work. 
Sec.~\ref{sec:methodology} introduces ChronoChart. 
Sec.~\ref{sec:experimental_setup} describes the datasets, baselines, and evaluation metrics. 
Sec.~\ref{sec:results} presents the results and ablation studies, before Sec.~\ref{sec:conclusion} concludes.

\section{Related Work}\label{sec:related_work}

\begin{table}[t]
\centering
\caption{Comparison of supervision and setup requirements.}
\label{tab:requirements}
\resizebox{\columnwidth}{!}{%
\begin{tabular}{lccccc}
\textbf{Method} 
& \textbf{Labels $\downarrow$} 
& \textbf{Setup metric $\downarrow$} 
& \textbf{$O(n^2)$ matrix $\downarrow$} 
& \textbf{Aux. info $\downarrow$} 
& \textbf{Mini-batch $\uparrow$} \\
\midrule
Fingerprinting~\cite{ref2,ref3} 
& yes & no & no & no & yes \\

Siamese w. ADP/CIRA~\cite{ref7,adp,geodesic} 
& no & yes & yes & no & yes \\

PDR-based CC~\cite{pdr} 
& no & yes & yes & yes & yes \\

Geodesic loss~\cite{uncertainty-aware} 
& no & yes & yes & yes & no \\

Temporal AE~\cite{ae-cc} 
& no & no & no & yes & yes \\

\textbf{ChronoChart} 
& \textbf{no} & \textbf{no} & \textbf{no} & \textbf{no} & \textbf{yes} \\
\end{tabular}%
}
\vspace{-0.6cm}
\end{table}

\subsection{Fingerprinting-Based Localization}
Fingerprinting-based localization learns a mapping from radio measurements, such as RSS or \gls{csi}, to physical positions using supervised training data. 
These methods achieve high accuracy when dense labeled fingerprints are available~\cite{ref2,ref3}. 
However, collecting reference positions requires labor-intensive site surveys, and changes in the environment degrade performance and require recalibration~\cite{ref2,ref3}. 
This motivates label-free alternatives that reduce deployment effort while preserving localization accuracy.

\subsection{Channel Charting}
\gls{cc} was introduced as a label-free paradigm that maps high-dimensional \gls{csi} to a low-dimensional chart reflecting radio geometry~\cite{ref4,ref6}. 
Early approaches extract slowly varying channel features, such as large-scale fading, and apply non-parametric dimensionality reduction methods such as PCA, MDS, Isomap, or Sammon mapping~\cite{ref4,ref13}. 
As \gls{cc} does not require ground-truth coordinates during training, it substantially reduces the labeling effort compared with supervised localization~\cite{ref4,ref12}. 
However, non-parametric methods do not provide a parametric mapping for unseen measurements, limiting their use in online localization.
Deep-learning-based \gls{cc} addresses this limitation by learning encoders that embed new samples. 
Autoencoders~\cite{ref4,ae-cc}, Siamese networks~\cite{ref7}, and triplet networks~\cite{triplet,rappaport2021improving} learn channel charts from radio measurements using pairwise or triplet constraints derived from the input data itself or from side information. 
Invariant and self-supervised representation learning improves robustness~\cite{ref12}. 
However, purely local or relative constraints still struggle to recover globally coherent charts without additional structure.
To improve global consistency, several methods introduce radio-specific dissimilarity metrics, geodesic distances, or additional localization constraints. 
Examples include angle-delay-profile-based and timestamp-aided dissimilarity metrics~\cite{adp}, CIRA-based global \gls{cc}~\cite{geodesic}, PDR-based \gls{cc}~\cite{pdr}, and batch-wise geodesic losses with acceleration constraints~\cite{uncertainty-aware}. 
Others~\cite{euchner2023augmenting,cc-real,uniloc,tdoa-cc} combine \gls{cc} with classical localization techniques or geometric constraints to obtain charts in real-world coordinates. 
Temporal or inertial constraints have also been used to encourage smooth trajectories in the learned representation~\cite{ae-cc,adp,rappaport2021improving,uncertainty-aware}. 
While effective, these methods rely on setup-specific metrics, auxiliary sensors, propagation or motion priors, graph construction parameters, or precomputed pairwise distance matrices. 
This improves accuracy, but limits scalability and transferability across heterogeneous radio systems.

\subsection{Self-Supervised Representation Learning}
Self-supervised representation learning learns embeddings without manual labels by exploiting structure in the data. 
Masked prediction methods randomly mask parts of the input and reconstruct the missing information from the remaining context~\cite{ref10}. 
Joint-embedding methods enforce consistency between representations of different views or augmentations of the same sample~\cite{pi-model,grill2020bootstrap}. 
To avoid representation collapse, some methods use exponential moving averages of a target network~\cite{grill2020bootstrap}, while consistency-based semi-supervised methods operate without such teacher models~\cite{pi-model}. 

However, image-style augmentations do not directly transfer to radio measurements, where suitable invariances are less obvious. So, ChronoChart uses masking as a radio-agnostic perturbation and enforces consistency between masked and unmasked \gls{csi} representations. 
This joint-embedding objective complements the temporal triplet and temporal-proximity losses by promoting robust, data-driven embeddings.

\section{Methodology}\label{sec:methodology}

ChronoChart uses only \gls{csi} and timestamps for training. 
It replaces setup-specific distance engineering with timestamp-supervised learning and does not require radio-specific dissimilarity metrics, inertial measurements, floor maps, or precomputed pairwise distance matrices.
While temporal constraints have been used before, existing approaches often rely on autoencoder reconstruction, handcrafted dissimilarity metrics, thresholded graph construction, or additional motion information. 
ChronoChart instead uses timestamps only to generate weak training relations and combines them with contrastive and consistency-based objectives in a mini-batch training framework.

This Sec. formulates the label-free channel charting problem (Sec.~\ref{subsec:problem}), introduces the key idea (Sec.~\ref{subsec:concept}), describes the architectures (Sec.~\ref{subsec:arch}), training objectives (Sec.~\ref{subsec:objectives}), and NashMTL-based optimization (Sec.~\ref{subsec:multi_objective}).

\subsection{Problem Formulation}\label{subsec:problem}

Let $\mathbf{H}_i \in \mathbb{C}^{A \times S}$ denote a \gls{csi} snapshot recorded at timestamp $t_i$, where $A$ is the number of antennas and $S$ is the number of subcarriers. 
The training set is given by
$\mathcal{D} = \{(\mathbf{H}_i,t_i)\}_{i=1}^{N},$
and contains no position labels. 
For evaluation only, let $\mathbf{p}_i \in \mathbb{R}^{D}$ denote the corresponding physical position.

The goal of \gls{cc} is to learn an encoder
$f_\theta:\mathbf{H}_i \mapsto \mathbf{z}_i \in \mathbb{R}^{d},$
that maps each \gls{csi} snapshot to a low-dimensional channel-chart representation $\mathbf{z}_i$, such that distances in the chart reflect spatial relationships in the physical environment. 
Ideally, samples that are nearby in physical space, $\|\mathbf{p}_i-\mathbf{p}_j\|_2$, should also be nearby in the chart, $\|\mathbf{z}_i-\mathbf{z}_j\|_2$.
Since the chart is learned without absolute position labels, its coordinate system is identifiable only up to a global transformation and is aligned to physical coordinates only for evaluation.

ChronoChart addresses this problem under a restricted supervision setting: during training, it uses only \gls{csi} samples and timestamps. 
It does not require position labels, radio-specific dissimilarity metrics, propagation priors, floor maps, inertial measurements, or precomputed pairwise distance matrices.

\subsection{Key Idea}\label{subsec:concept}

ChronoChart uses the timestamps in $\mathcal{D}=\{(\mathbf{H}_i,t_i)\}_{i=1}^{N}$ to derive weak spatial relations between \gls{csi} samples. 
For an anchor sample $\mathbf{H}_i$, temporally close samples are treated as positive candidates, while samples outside this temporal neighborhood serve as weak negative candidates~\cite{triplet}:
$\mathcal{P}_i=\{j\neq i \mid |t_i-t_j|\leq T\}$ and $\mathcal{N}_i=\{j\neq i \mid |t_i-t_j|>T\}$.
This exploits the assumption that, under continuous motion, small temporal differences usually imply small spatial displacements.

As temporal proximity is only a weak cue and is imperfect under non-uniform motion, trajectory crossings, or near-stationary periods, ChronoChart combines timestamp-derived supervision with representation consistency. 
Given embeddings $\mathbf{z}_i=f_\theta(\mathbf{H}_i)$, the training objective consists of three complementary terms: a timestamp-based triplet loss that enforces relative ordering between positives and negatives, a temporal-proximity loss that regularizes distances between temporally adjacent embeddings, and a joint-embedding loss that aligns two views of the same \gls{csi} sample. 
The resulting multi-objective problem is optimized with NashMTL to reduce conflicting gradients between the objectives.

\subsection{Model Architecture}\label{subsec:arch}

ChronoChart uses a shared encoder $f_\theta$ to map \gls{csi} measurements to low-dimensional channel-chart embeddings.\footnote{We use the same encoder architectures as the baseline methods for each dataset, to ensure that performance gains stem from ChronoChart.} 

\subsubsection{MLP-Based Model}
For the Dichasus dataset, we use a \gls{mlp}-based~\cite{adp} architecture. 
The input features are cross-correlations computed between \glspl{cir} from all available antennas. 
Unlike the baseline model, that uses batch normalization, we adopt layer normalization to avoid dependence on batch-level statistics, that is unstable for small batch sizes.

\subsubsection{CNN-Based Model}
For the Fraunhofer 5G and passive UWB datasets, we use a CNN-based~\cite{geodesic} architecture. 
We apply normalization independently of batch statistics, using InstanceNorm2D in the convolutional layers and layer normalization in the \gls{mlp} components.

\subsection{Training Objectives}\label{subsec:objectives}
\subsubsection{Timestamp-Based Triplet Loss}
The triplet loss provides the structural backbone of ChronoChart. 
For each anchor sample $\mathbf{H}_a$, we sample a positive $\mathbf{H}_p$ from its temporal neighborhood and a negative $\mathbf{H}_n$ from outside this neighborhood, as defined in Sec.~\ref{subsec:concept}. 
The corresponding embeddings are $\mathbf{z}_a=f_\theta(\mathbf{H}_a)$, $\mathbf{z}_p=f_\theta(\mathbf{H}_p)$, and $\mathbf{z}_n=f_\theta(\mathbf{H}_n)$. 
We define
\[
d_{\mathrm{pos}}=\|\mathbf{z}_a-\mathbf{z}_p\|_2,
\qquad
d_{\mathrm{neg}}=\|\mathbf{z}_a-\mathbf{z}_n\|_2 .
\]
The softmax-based triplet loss is
\begin{equation}
\mathcal{L}_{\mathrm{triplet}}
=
-\log
\left(
\frac{\exp(-d_{\mathrm{pos}})}
{\exp(-d_{\mathrm{pos}})+\exp(-d_{\mathrm{neg}})}
\right).
\end{equation}
This objective encourages temporally close samples to be embedded closer than temporally distant samples. 
Unlike distance-matrix-based Siamese methods, the triplet loss only requires sampled temporal relations and therefore is optimized on standard mini-batches. 
It prevents representation collapse and establishes local as well as coarse global chart structure. 
Positive samples are selected within a temporal window of length $T$\footnote{$T$ for Dichasus: 2.5s, 5G: 1.0s, and Passive UWB: 1.9s. Results remain stable ($<$18\% relative variation) for $T$ within $\sim$2s of the chosen value.}, while negatives are sampled from outside this window. 
The temporal window $T$ is the only sampling parameter used to define positive candidates. 
It does not encode a radio-specific distance metric and is not used to construct a global pairwise distance matrix.

\subsubsection{Temporal-Proximity Loss}
The triplet loss enforces relative ordering between positives and negatives, but does not explicitly constrain the distance between temporally adjacent samples. 
We therefore add a temporal-proximity loss that encourages smooth trajectories in the channel chart. 
Let $\mathcal{T}$ denote the set of anchor-positive pairs in a mini-batch and let $\Delta t_{ij}=|t_i-t_j|$ be the temporal distance between samples $i$ and $j$. 
We assign each pair the weight
$w_{ij} =1-\frac{2\,\Delta t_{ij}}{T},$
so that closer temporal neighbors receive stronger emphasis.
The temporal-proximity loss is defined as
\begin{equation}
\mathcal{L}_{\mathrm{time}}
=
\frac{
\sum_{(i,j)\in\mathcal{T}}
w_{ij}
\left(
\|\mathbf{z}_i-\mathbf{z}_j\|_2-k\Delta t_{ij}
\right)^2
}{
\sum_{(i,j)\in\mathcal{T}} w_{ij}
}.
\end{equation}
where \(k > 0\) is a trainable scaling factor that aligns temporal distances with the scale of the embedding space, ensuring that temporal differences and embedding distances are comparable during optimization.
This equation regularizes the chart trajectory by encouraging embeddings of temporally adjacent samples to remain close. 
As temporal distance is only a weak proxy for spatial displacement, especially under non-uniform motion, this loss is used as a complementary constraint rather than as a standalone supervision signal.

\begin{figure*}[t]
    \centering
    \begin{subfigure}[t]{0.19\linewidth}
        \centering
        \includegraphics[width=\linewidth, height=3cm, keepaspectratio]{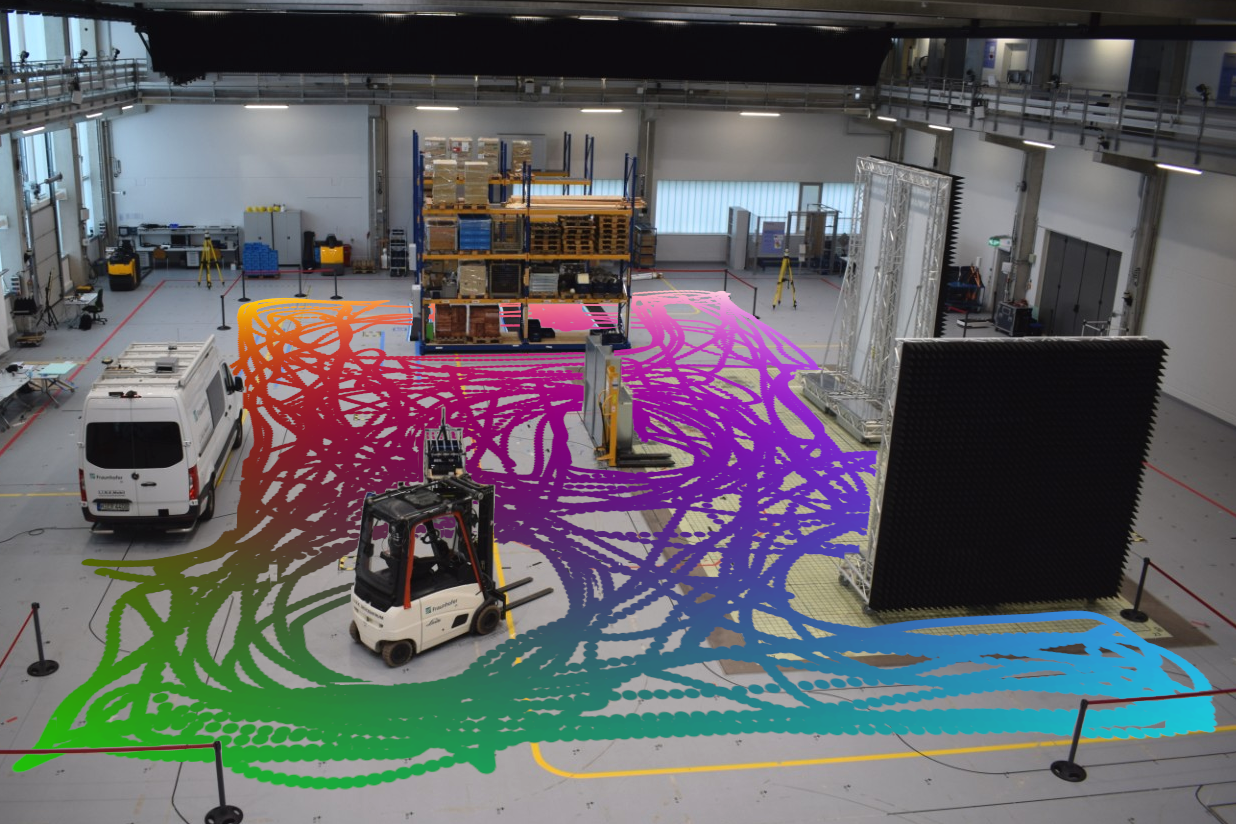}
        \caption{5G Reference}
    \end{subfigure}\hfill
    \foreach \name/\cap in {5g_pca/5G PCA, 5g_isomap/5G Isomap, 5g_siamese_cira/5G Siamese, 5g_ours/5G ChronoChart}{%
        \begin{subfigure}[t]{0.19\linewidth}
            \centering
            \includegraphics[width=\linewidth, height=3cm, keepaspectratio]{fig/subplots/\name.pdf}
            \caption{\cap}
        \end{subfigure}\hfill
    }

    \vspace{0.1cm}

    \begin{subfigure}[t]{0.19\linewidth}
        \centering
        \includegraphics[width=\linewidth, height=2.4cm, keepaspectratio]{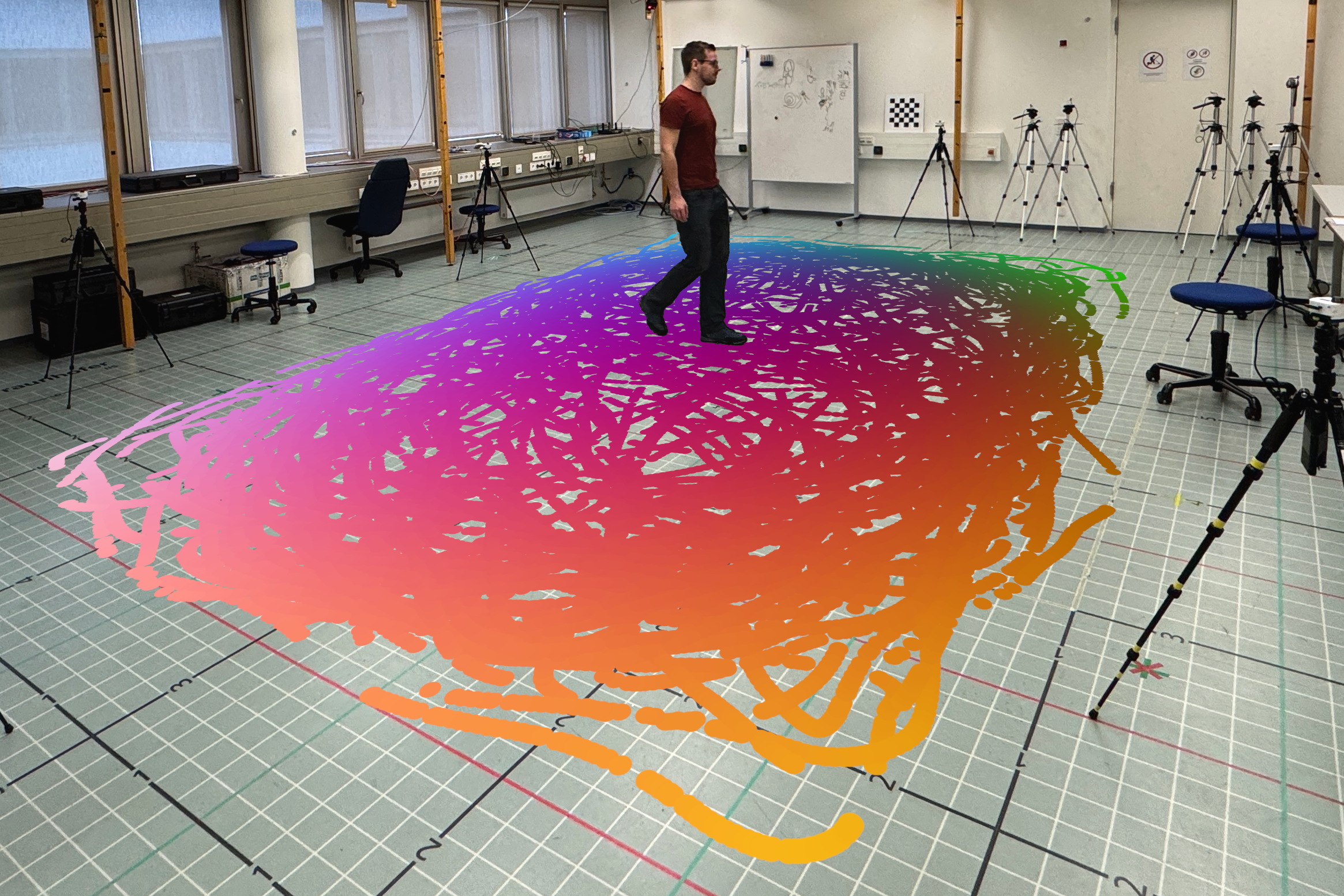}
        \caption{Passive Reference}
    \end{subfigure}\hfill
    \foreach \name/\cap in {passive_pca/Passive PCA, passive_isomap/Passive Isomap, passive_siamese_pdr/Passive Siamese PDR, passive_ours/Passive ChronoChart}{%
        \begin{subfigure}[t]{0.19\linewidth}
            \centering
            \includegraphics[width=\linewidth, height=2.4cm, keepaspectratio]{fig/subplots/\name.pdf}
            \caption{\cap}
        \end{subfigure}\hfill
    }
    \caption{Qualitative comparison of channel charts for the Fraunhofer 5G and passive UWB datasets.}
    \label{fig:results_part1}
    \vspace{-0.4cm}
\end{figure*}

\subsubsection{Joint-Embedding Loss}
Temporal proximity alone is unreliable under non-uniform motion, trajectory crossings, or near-stationary periods. 
To complement the timestamp-derived objectives, we introduce a joint-embedding loss that promotes robust, data-driven representations. 
Following consistency-based self-supervised learning~\cite{pi-model}, we leverage two complementary views of the same input: the original, unmodified features and a masked variant.
Let $\mathbf{z}_i=f_\theta(\mathbf{H}_i)$ denote the embedding of the unmasked input and let $\widehat{\mathbf{z}}_i$ denote the embedding obtained when a stochastic feature mask is applied. 
We use a masking ratio of $90\%$. 
For the \gls{mlp}-based architecture, we mask after the cross-correlation features are computed. 
For the CNN-based architecture, we mask after the convolutional feature extractor and before the regression \gls{mlp}. 
The joint-embedding loss is
\begin{equation}
\mathcal{L}_{\mathrm{JE}}
=
\frac{1}{|\mathcal{B}|}
\sum_{i\in\mathcal{B}}
\|\mathbf{z}_i-\widehat{\mathbf{z}}_i\|_2^2 ,
\end{equation}
where $\mathcal{B}$ denotes the mini-batch.
This objective enforces consistency between (un)masked representations, making the embedding less sensitive to missing, noisy, or corrupted input features. 
It encourages the encoder to capture stable channel structure that is not solely determined by temporal adjacency.

\subsection{Multi-Objective Optimization}\label{subsec:multi_objective}

ChronoChart combines three objectives: the timestamp-based triplet loss $\mathcal{L}_{\mathrm{triplet}}$, the temporal-proximity loss $\mathcal{L}_{\mathrm{time}}$, and the joint-embedding loss $\mathcal{L}_{\mathrm{JE}}$. In a multi-objective setup, gradients from different losses conflict, which may cause suboptimal convergence or training oscillations. Instead of combining the losses with fixed static weights, we use NashMTL to compute a conflict-reduced update direction at each training step. Let $\mathbf{g}_k = \nabla_\theta \mathcal{L}_k$ denote the gradient of objective $k$. NashMTL formulates multi-task optimization as a bargaining game, solving for a Nash equilibrium among objectives at each step to derive adaptive per-task weights $\alpha_k$. The final update direction is $\sum_k \alpha_k \mathbf{g}_k$, where the $\alpha_k$ are obtained by solving a constrained quadratic program. This reduces destructive interference between objectives while avoiding manually tuned loss weights.
For the ablation study, we use UPGrad~\cite{upgrad}, which exposes interpretable coefficients $\lambda_1$, $\lambda_2$, and $\lambda_3$ for the triplet, temporal-proximity, and joint-embedding losses, respectively. 
This allows us to study the contribution of each component by varying $\lambda_2$ and $\lambda_3$ while keeping $\lambda_1=1$ fixed.

\begin{table}[t]
\centering
\caption{Statistics [m/s]; coefficient of variation (CoV).}
\label{tab:motion_stats}
\resizebox{0.8\columnwidth}{!}{%
\begin{tabular}{lccccc}
\textbf{Dataset} & \textbf{Mean} & \textbf{Std} & \textbf{Min} & \textbf{Max} & \textbf{CoV} \\
\midrule
Dichasus & 0.227 & 0.082 & 0.000 & 0.729 & 0.362 \\
Fraunhofer 5G & 0.902 & 0.351 & 0.000 & 2.295 & 0.389 \\
Passive UWB & 0.703 & 0.135 & 0.137 & 0.980 & 0.193 \\
\end{tabular}
}
\vspace{-0.5cm}
\end{table}

\section{Experimental Setup}\label{sec:experimental_setup}

\subsection{Datasets}

We evaluate ChronoChart on three real-world indoor datasets covering heterogeneous radio setups: 5G, distributed MIMO, and passive UWB.
For all datasets, ChronoChart is trained only on radio measurements and timestamps without position labels. 
Ground-truth positions are used only for evaluation, where the learned chart is aligned to physical coordinates using an optimal affine transformation. 
To characterize the motion conditions, Tbl.~\ref{tab:motion_stats} reports velocity statistics derived from the ground-truth trajectories for each dataset. 
\noindent\textit{Fraunhofer 5G dataset~\cite{fraunhofer_5g}} contains \gls{csi} measurements collected from a 5G base station with six distributed antennas in an industrial indoor environment with rich multipath propagation. 
The setup represents an active radio localization scenario with a {SISO} transmitter and distributed receiver infrastructure.
\noindent\textit{Dichasus dataset~\cite{dichasus2021}} contains distributed {MIMO} channel measurements acquired with 4 antenna arrays, each equipped with 8 antennas, placed around an L-shaped indoor area with multipath propagation. 
\noindent\textit{Passive UWB dataset~\cite{poeggel2025passive}} contains approximately 40 minutes of training data and two test sets, denoted P1 and P2 of a passive UWB system running in an office-like environment using a fully connected 14-link UWB {mesh} and three human subjects.

\subsection{Training Protocol}
We train all models using the Adam optimizer with a learning rate of $10^{-3}$, a batch size of 512, and 200 training epochs. 
As training is stochastic, each experiment is repeated 15 times with different random seeds, and we report the median performance. 
ChronoChart uses the same encoder architecture as the corresponding baseline for each dataset, while changing only the training objective.

\subsection{Baseline Methods}
We compare ChronoChart against non-parametric, neural, and metric-based \gls{cc} baselines.
{PCA and Isomap:} Classical non-parametric dimensionality reduction methods applied to the \gls{csi} features~\cite{ref13}. They provide label-free embeddings but do not learn an encoder for unseen samples. {Triplet-Only CC:} A neural \gls{cc} baseline trained only with the timestamp-based triplet loss~\cite{triplet}. This baseline isolates the effect of the temporal-proximity and joint-embedding losses. {Siamese Network:} A pairwise network trained with setup-specific distance constraints~\cite{ref7}. We use ADP-based distances for Dichasus~\cite{adp}, CIRA-based distances for the Fraunhofer 5G dataset~\cite{geodesic}, and PDR-based distances for the passive UWB dataset~\cite{pdr}. 
{Geodesic Loss:} For the Dichasus dataset, we additionally report the uncertainty-aware geodesic loss method with acceleration constraints from~\cite{uncertainty-aware}.

\subsection{Evaluation Metrics}
We evaluate localization accuracy after aligning the learned chart to ground-truth coordinates using an optimal affine transformation. 
Let $\hat{\mathbf{p}}_i$ be the aligned estimate and $e_i=\|\hat{\mathbf{p}}_i-\mathbf{p}_i\|_2$ the localization error. 
We report $\mathrm{MAE}=\frac{1}{N}\sum_i e_i$, $\mathrm{DRMS}=\sqrt{\frac{1}{N}\sum_i e_i^2}$, the median error $\mathrm{CEP}$, and the 95th-percentile error $\mathrm{R95}$, all in meters. 
For chart quality, we report continuity (CT), trustworthiness (TW), and Kruskal stress (KS). 
Higher CT/TW indicate better local neighborhood preservation; lower KS indicates better global distance preservation.

\begin{table}[!t]
\centering
\caption{Performance comparison of different methods}
\label{tab:performance_comparison_all}
\resizebox{\columnwidth}{!}{%
\begin{tabular}{llcccccc}
\textbf{Data} & \textbf{Method} & \textbf{MAE $\downarrow$} & \textbf{DRMS $\downarrow$} & \textbf{CEP $\downarrow$} & \textbf{R95 $\downarrow$} & \textbf{KS $\downarrow$} & \textbf{CT/TW $\uparrow$} \\
\hline
\multirow{6}{*}{\rotatebox[origin=c]{90}{\textit{Dichasus}}}
 & PCA                        & 2.915\,m & 3.294\,m & 2.786\,m & 5.670\,m & 0.567 & 0.836 / 0.805 \\
 & Isomap                     & 2.842\,m & 3.216\,m & 2.753\,m & 5.368\,m & 0.537 & 0.818 / 0.810 \\
 & Triplet-only               & 1.230\,m & 1.554\,m & 0.907\,m & 3.251\,m & 0.218 & 0.987 / 0.973 \\
 & Siamese w. ADP~\cite{adp}& 0.490\,m & 0.584\,m & 0.441\,m & 1.026\,m & 0.071 & 0.996 / 0.996 \\
 & Geodesic loss~\cite{uncertainty-aware} & 0.330\,m & 0.409\,m & 0.277\,m & 0.719\,m & 0.069 & 0.998 / 0.998 \\
 & \textbf{ChronoChart}   & \textbf{0.264\,m} & \textbf{0.329\,m} & \textbf{0.217\,m} & \textbf{0.633\,m} & \textbf{0.0546} & \textbf{0.998 / 0.998} \\
\hline
\multirow{5}{*}{\rotatebox[origin=c]{90}{\textit{\shortstack{Fraunhofer\\5G}}}}
& PCA                        & 3.646\,m & 4.116\,m & 3.383\,m & 7.143\,m & 0.327 & 0.919 / 0.902 \\
 & Isomap                     & 2.696\,m & 3.164\,m & 2.394\,m & 5.932\,m & 0.246 & 0.961 / 0.949 \\
 & Triplet-only               & 1.453\,m & 1.842\,m & 1.219\,m & 3.202\,m & 0.139 & 0.987 / 0.987 \\
 & Siamese w. CIRA~\cite{geodesic} & 1.331\,m & 1.601\,m & 1.187\,m & 2.790\,m & 0.139 & 0.985 / 0.984 \\
 & \textbf{ChronoChart}   & \textbf{0.692\,m} & \textbf{1.232\,m} & \textbf{0.419\,m} & \textbf{1.733\,m} & \textbf{0.087} & \textbf{0.991 / 0.995} \\
\hline
\multirow{5}{*}{\rotatebox[origin=c]{90}{\textit{\shortstack{Passive\\UWB P1}}}}
 & PCA                        & 1.335\,m & 1.462\,m & 1.344\,m & 2.262\,m & 0.65  & 0.626 / 0.584 \\
 & Isomap                     & 1.436\,m & 1.536\,m & 1.490\,m & 2.231\,m & 0.699 & 0.560 / 0.548 \\
 & Triplet only               & 0.241\,m & 0.287\,m & 0.214\,m & 0.514\,m & 0.139 & 0.983 / 0.984 \\
 & Siamese w. PDR~\cite{poeggel2025passive} & 0.273\,m & 0.326\,m & 0.238\,m & 0.587\,m & 0.145 & 0.974 / 0.974 \\
 & \textbf{ChronoChart}   & \textbf{0.164\,m} & \textbf{0.198\,m} & \textbf{0.144\,m} & \textbf{0.349\,m} & \textbf{0.095} & \textbf{0.993 / 0.993} \\
\hline
\multirow{5}{*}{\rotatebox[origin=c]{90}{\textit{\shortstack{Passive\\UWB P2}}}} & PCA                        & 1.502\,m & 1.617\,m & 1.580\,m & 2.357\,m & 0.688 & 0.606 / 0.585 \\
 & Isomap                     & 1.546\,m & 1.648\,m & 1.647\,m & 2.233\,m & 0.742 & 0.543 / 0.527 \\
 & Triplet only               & 0.280\,m & 0.324\,m & 0.257\,m & 0.572\,m & 0.141 & 0.982 / 0.983 \\
 & Siamese w. PDR~\cite{poeggel2025passive} & 0.236\,m & 0.287\,m & 0.204\,m & 0.508\,m & 0.127 & 0.981 / 0.981 \\
 & \textbf{ChronoChart}   & \textbf{0.195\,m} & \textbf{0.228\,m} & \textbf{0.178\,m} & \textbf{0.393\,m} & \textbf{0.096} & \textbf{0.991 / 0.991} \\
\end{tabular}
}
\vspace{-0.5cm}
\end{table}

\section{Results}\label{sec:results}

\subsection{Overall Performance}
Tbl.~\ref{tab:performance_comparison_all} compares ChronoChart with non-parametric, triplet-only, Siamese, and geodesic-loss baselines across all datasets. 
ChronoChart consistently improves over PCA, Isomap, and triplet-only training. 
On the Fraunhofer 5G and passive UWB datasets, ChronoChart achieves the best localization performance among the compared methods. 
For example, on Fraunhofer 5G, it reduces the MAE from $1.331\,\mathrm{m}$ for the CIRA-based Siamese baseline to $0.692\,\mathrm{m}$. 
On passive UWB, ChronoChart also outperforms the PDR-based Siamese baseline on both P1 and P2.
On Dichasus, ChronoChart marginally achieves the lowest error, with an MAE approximately 7 cm below the geodesic-loss method. 
However, this method uses geodesic constraints and acceleration regularization, whereas ChronoChart uses only \gls{csi} and timestamps during training. 
Thus, ChronoChart achieves competitive accuracy while avoiding setup-specific distance engineering.
The qualitative results in Fig.~\ref{fig:results_part1} support the quantitative findings. 
ChronoChart produces channel charts with coherent global structure and smooth trajectories across different radio modalities. 
The charts show that timestamp-supervised training recovers meaningful spatial structure without relying on precomputed pairwise distance matrices or radio-specific dissimilarity metrics.

\subsection{Scalability and Setup-Agnosticity}
The main advantage of ChronoChart is not only localization accuracy, but also the reduced amount of setup-specific engineering. 
Siamese and geodesic baselines require distance metrics tailored to the radio system, such as ADP for Dichasus, CIRA for 5G, or PDR-based distances for passive UWB. 
In addition, geodesic methods require precomputing and storing pairwise distance matrices with $O(n^2)$ memory complexity. 
The geodesic-loss method further requires processing the full dataset jointly to enforce acceleration constraints. 
In contrast, ChronoChart operates on standard mini-batches, requires no precomputed distance matrix, and uses the same timestamp-supervised training principle across all datasets.
Beyond standard training hyperparameters, ChronoChart introduces only a temporal sampling window $T$ and a masking ratio. 
The window $T$ is used to sample weak positive candidates and does not define a radio-specific distance metric or graph structure. 
This distinguishes us from methods that require manually designed dissimilarity metrics, graph construction parameters, auxiliary motion information, or environment-specific priors.

\subsection{Ablation Study}

Tbl.~\ref{tab:loss_coefficients} studies the contribution of the temporal-proximity and joint-embedding losses using UPGrad coefficients $\lambda_2$ and $\lambda_3$, while keeping the triplet coefficient fixed at $\lambda_1=1$. 
The entry $(\lambda_2=0,\lambda_3=0)$ corresponds to the triplet-only baseline. 
On both Fraunhofer 5G and Dichasus, adding either the temporal-proximity loss or the joint-embedding loss improves performance substantially, showing that both objectives add information beyond the triplet loss.
The best results are obtained when both losses are active. 
For Fraunhofer 5G, the error decreases from $2.18\,\mathrm{m}$ for triplet-only training to $0.84\,\mathrm{m}$ when both objectives are used. 
For Dichasus, the error decreases from $1.23\,\mathrm{m}$ to $0.38\,\mathrm{m}$. 
Performance remains relatively stable across a range of nonzero $\lambda_2$ and $\lambda_3$ values, indicating that the method is not overly sensitive to the exact weighting of the auxiliary objectives.
These results support the intended roles of the three objectives. 
The triplet loss provides the structural backbone and prevents collapse, the temporal-proximity loss promotes smooth chart trajectories, and the joint-embedding loss improves robustness by enforcing consistency between masked and unmasked representations. 
The improvements persist despite non-uniform motion (Tbl.~\ref{tab:motion_stats}), suggesting that ChronoChart is robust to realistic trajectory variations and near-stationary periods.

\begin{table}[t]
\centering
\caption{Comparison of different UPGrad loss coefficients for the temporal loss ($\lambda_2$) and the joint-embedding loss ($\lambda_3$). Values denote Euclidean localization error in meters.}
\vspace{-0.2cm}
\label{tab:loss_coefficients}
\setlength{\tabcolsep}{4pt}
\definecolor{lightgraytable}{gray}{0.97}
\resizebox{0.7\columnwidth}{!}{%
\begin{tabular}{rrrrrp{0.0em}rrrrr}
\multicolumn{5}{c}{\textbf{Fraunhofer 5G Dataset}} 
& 
& \multicolumn{5}{c}{\textbf{Dichasus Dataset}} \\
\cmidrule(lr){1-5} \cmidrule(lr){7-11}

& \multicolumn{4}{c}{Temp. loss coeff. $\lambda_2$}
& &
& \multicolumn{4}{c}{Temp. loss coeff. $\lambda_2$} \\
\cmidrule(lr){2-5} \cmidrule(lr){8-11}

\cellcolor{lightgraytable}$\lambda_3$
& \multicolumn{1}{c}{\cellcolor{lightgraytable}0.0}
& \multicolumn{1}{c}{\cellcolor{lightgraytable}0.5}
& \multicolumn{1}{c}{\cellcolor{lightgraytable}1.0}
& \multicolumn{1}{c}{\cellcolor{lightgraytable}1.5}
&
& \cellcolor{lightgraytable}$\lambda_3$
& \multicolumn{1}{c}{\cellcolor{lightgraytable}0.0}
& \multicolumn{1}{c}{\cellcolor{lightgraytable}0.5}
& \multicolumn{1}{c}{\cellcolor{lightgraytable}1.0}
& \multicolumn{1}{c}{\cellcolor{lightgraytable}1.5} \\
\cmidrule(){1-5} \cmidrule(){7-11}

\cellcolor{lightgraytable}0.0 & 2.18 & 0.99 & 1.04 & 0.97
& & \cellcolor{lightgraytable}0.0 & 1.23 & 0.62 & 0.80 & 0.55 \\

\cellcolor{lightgraytable}0.5 & 1.01 & 0.94 & 0.98 & 0.98
& & \cellcolor{lightgraytable}0.5 & 0.66 & 0.40 & 0.48 & 0.40 \\

\cellcolor{lightgraytable}1.0 & 1.01 & \textbf{0.84} & 0.98 & 0.94
& & \cellcolor{lightgraytable}1.0 & 0.48 & \textbf{0.38} & \textbf{0.38} & 0.39 \\

\cellcolor{lightgraytable}1.5 & 1.14 & 0.86 & 0.98 & 0.86
& & \cellcolor{lightgraytable}1.5 & 0.66 & 0.41 & 0.39 & 0.39 \\
\end{tabular}
}
\vspace{-0.5cm}
\end{table}

\section{Conclusion}\label{sec:conclusion}

We presented ChronoChart, a timestamp-supervised \gls{cc} framework for label-free indoor localization. 
It learns channel charts from \gls{csi} samples and timestamps only, without requiring priors, i.e., position labels, radio-specific dissimilarity metrics, propagation priors, floor maps, inertial measurements, or precomputed pairwise distance matrices. 
The framework combines three complementary objectives: a timestamp-based triplet loss for coarse chart structure, a temporal-proximity loss for trajectory smoothness, and a joint-embedding loss for robust representations. 
Experiments on real-world 5G, distributed MIMO, and passive UWB datasets show that ChronoChart matches or outperforms baselines that rely on tailored distance metrics, while remaining scalable and applicable across heterogeneous radio systems. 
Ablation studies confirm that the 3 objectives provide complementary improvements.
As ChronoChart uses temporal proximity as a weak proxy for spatial proximity it fits continuously moving transmitters or objects, but may be violated under long stationary periods, irregular sampling, or highly non-uniform motion. 

\vspace{-0.2cm}

\AtNextBibliography{\footnotesize}
\printbibliography

\end{document}